\documentclass[aps,prb,letterpaper,amsmath,amssymb,superscriptaddress,nofootinbib,twocolumn]{revtex4-2}

\usepackage{graphicx}
\usepackage[usenames]{color}
\usepackage[colorlinks=true,linkcolor=blue,citecolor=blue,anchorcolor=green,urlcolor=blue]{hyperref}
\usepackage{orcidlink}

\usepackage[percent]{overpic}
\usepackage{tabularx}
\usepackage{multirow}

\usepackage{CJK}
\renewcommand{\selectlanguage}[1]{}

\def\be{\begin{equation}}
\def\ee{\end{equation}}
\def\bea{\begin{eqnarray}}
\def\eea{\end{eqnarray}}

\begin{document}

\begin{CJK*}{UTF8}{}
\title{
Truncated string state space approach and its application to nonintegrable spin-\texorpdfstring{$\frac{1}{2}$}{} Heisenberg chain
}

\author{Jiahao~Yang~(\CJKfamily{gbsn}杨家豪)\orcidlink{0000-0001-7670-2218}}
\affiliation{Tsung-Dao Lee Institute, Shanghai Jiao Tong University, Shanghai 201210, China}

\author{Jianda Wu (\CJKfamily{gbsn}吴建达)\orcidlink{0000-0002-3571-3348}}
\email{wujd@sjtu.edu.cn}
\affiliation{Tsung-Dao Lee Institute, Shanghai Jiao Tong University, Shanghai 201210, China}
\affiliation{School of Physics and Astronomy, Shanghai Jiao Tong University, Shanghai 200240, China}
\affiliation{Shanghai Branch, Hefei National Laboratory, Shanghai 201315, China}

\begin{abstract}
By circumventing the difficulty of obtaining exact string state solutions to Bethe ansatz equations,
we devise a truncated string state space approach for investigating spin dynamics in a nonintegrable spin-$\frac{1}{2}$ Heisenberg chain subjected to a staggered field at various magnetizations.
The obtained dynamical spectra reveal a series of elastic peaks at integer multiples of the ordering wave vector $Q$,
indicating the presence of multi-$Q$ Bethe string states within the ground state.
The spectrum exhibits a separation between different string continua as the strength of the staggered field increases at low magnetization,
reflecting the confinement of the Bethe strings.
This approach provides a unified string-state-based framework
for understanding spin dynamics in low-dimensional nonintegrable Heisenberg models,
which has a successful application to observations across various phases of the quasi-one-dimensional antiferromagnet $\rm YbAlO_3$.
\end{abstract}
\date{\today}
\maketitle
\end{CJK*}

\section{Introduction}
One-dimensional quantum systems characterized by exact solutions and quantum integrability
offer a fascinating arena to study many-body physics.
Notable examples include the one-dimensional (1D) spin-$\frac{1}{2}$ XXZ model \cite{Jimbo1995,Franchini2017,yang_magnetic_2023,He_quantum_2017},
Gaudin-Yang model \cite{GAUDIN1967,CNYang_1967,Guan_Fermi_2013},
Lieb-Liniger model \cite{LL_I_1963,LL_II_1963},
and quantum Ising models \cite{Pfeuty1970,Niemeijer_1967,2isingboson,a_b_zamolodchikov_integrals_1989}.
Although these models have paved the way for determining the eigenstates and eigenenergies of those systems,
it has long been a challenge to calculate their form factors and thus dynamical response, which was partly tackled recently
\cite{Li_2023,Guan_Fermi_2013,guan_quench_2018,kitanine_form_1999,kitanine_correlation_2000,yang_local_2022,iorgov_spin_2011,mussardo_statistical_2020}.
Empowered by the theoretical development,
the spin dynamics of celebrated many-body quasiparticles,
such as spinons \cite{Caux_2006,caux_two-spinon_2008,castillo_exact_2020}, strings \cite{takahashi_one-dimensional_1971,kohno_string_dynamically_2009,yang_string_1D_2019},
$E_8$ \cite{coldea_quantum_2010,jianda_E8_2014,xiao_cascade_2021,Zou_E8_2021,wang2023spin} and $D_8^{(1)}$ particles \cite{gao2024spin,xi2024emergent,coupleCFT},
have been extensively explored, providing crucial guidance for the experimental observations in quasi-1D materials \cite{yang_magnetic_2023,Lake_Multispinon_2013,wang_experimental_2018,bera_dispersions_2020,wang_quantum_2019,zhang_e8_observation_2020,amelin_e8_experimental_2020,Zou_E8_2021,wang2023spin}.
This progress has led to the cooperative effort of both theorists and experimentalists to unveil the intricate nature of these exotic phenomena.

Bearing real materials in mind, it becomes crucial to ask how robust integrable physics is against nonintegrable perturbations that may partially or fully break the conservation laws of integrable systems.
This has inspired extensive research focused on nonintegrable models
such as spin-$\frac{1}{2}$ ladders
\cite{Steinigeweg_ladder,batchelor_integrable_2007},
chains with a staggered magnetic field \cite{Essler_q1DHeisenberg_1997,Starykh_q1DSDW_2014},
frustrated spin chains \cite{Bonca_direct_1994},
and dimerized spin chains \cite{Meisner_zero_2003,Keselman_Dynamical_2020}.
Most studies are performed by using effective field theory \cite{Essler_q1DHeisenberg_1997,Starykh_q1DSDW_2014} 
or numerical methods such as the exact diagonalization (ED) \cite{Bonca_direct_1994,Meisner_zero_2003},
matrix product state \cite{Keselman_Dynamical_2020},
and quantum Monte Carlo methods \cite{Zhou_Amplitude_2021}.
However, on the one hand, numerical methods in general lack a clear understanding
of the essential physical picture, on the other hand, the effective field theory
can provide only limited insight within the low-energy and long-wavelength limit.
Therefore, a method able to go beyond those limitations is always desired.

At first glance it may seem promising to apply Bethe states to study nonintegrable systems.
However, a notorious open problem persists:
finding the precise {\it complex} solutions of the Bethe ansatz equation (BAE) for string states
\cite{takahashi_one-dimensional_1971,takahashi_one-dimensional_1972,fujita_large-n_2003,ilakovac_violation_1999,isler_violations_1993,essler_fine_1992}.
In the past few decades,
many approaches have been explored,
including a carefully-designed iterative method \cite{hagemans_deformed_2007} and a rational $Q$-system method \cite{marboe_fast_2017,bajnok_generalized_2020,hou_spin-s_2024}.
The former can easily access large system size
but suffers from many unphysical solutions with repeated roots.
Although the latter can solve the BAE for all exact solutions simultaneously, it is limited to a small system size.
Those shortcomings impede the practical application of Bethe states to understand nonintegrable systems of reasonable size.

In this paper, we first outline a solving machine for the spin-$\frac{1}{2}$ Heisenberg chain,
which can obtain exact solutions for Bethe string states.
Based on these string states,
we develop a truncated string-state-space approach (TS${}^3$A) to study nonintegrable Hamiltonian, specifically the spin-$\frac{1}{2}$ Heisenberg chain with a staggered field.
The TS${}^3$A can determine the eigenstate and eigenenergy for the nonintegrable Hamiltonian in the truncated Hilbert space.
Additionally, we evaluate its efficiency for small systems under various truncation schemes, involving different energy cutoffs and string lengths,
by comparing its performance to that of ED calculations.

Following the TS${}^3$A,
we analyze the nonintegrable spin dynamics in the spin-$\frac{1}{2}$ Heisenberg chain with staggered field characterized by wavevector $Q$.
In addition to the $Q$-ordering of the system,
a series of elastic peaks appear at $nQ\ (|n|=2,3,\ldots)$,
indicating the ground state contains multi-$Q$ Bethe string states.
Moreover, the staggered field plays the role of the confining field for the Bethe string states,
constraining the motion of spins along the chain.
The confinement of Bethe strings results in two separated continua in dynamical spectra at low magnetization.
Notably, the above results were successfully applied to experimental observations of the quasi-1D antiferromagnet $\rm YbAlO_3$,
aligning with the unified Bethe-string-based framework provided by the TS${}^3$A
\cite{Yang_CBS_2023}.

The rest of this paper is organized as follows.
Section \ref{sec:model} introduces the Hamiltonian of the 1D Heisenberg model with a staggered field.
Section~\ref{sec:Bethe string state} illustrates the Bethe string state
and then presents a method for obtaining exact solutions from the BAE.
The framework of the TS$^3$A is developed, and its efficiency is investigated in Sec.~\ref{sec:TS3A}.
Then Sec.~\ref{sec:spin dynamics}  discusses the spin dynamics of the nonintegrable Hamiltonian.
Section~\ref{sec:conclusion} contains the conclusion and outlook.

\section{Model}
\label{sec:model}

Our parent Hamiltonian is the 1D Heisenberg spin-$\frac{1}{2}$ model with longitudinal field $h_z$,
\be
H_0= J \sum_{n=1}^N \mathbf{S}_n \cdot\mathbf{S}_{n+1}
    - h_z S_n^z
\label{eq:H0}
\ee
with $N$ being the total number of sites, $J$ being antiferromagnetic coupling,
and $\mathbf{S}_n$ being spin operators with components $S_n^\mu$ ($\mu=x,y,z$) at site $n$.
With the introduction of a staggered field $\mathbf{h}_Q=h_{Q}\sum_{i}\cos(Qr_i)\hat{z}$ which couples to the spin chain,
the total Hamiltonian becomes nonintegrable,
\be
H = H_0 + H',
\label{eq:H}
\ee
where
\be
H'=-\sum_n\mathbf{h}_Q\cdot \mathbf{S}_n= - h_{Q} \sum_n \cos(Q r_n) S^z_n.
\label{eq:Hprime}
\ee
$h_{Q}$ is the strength of the staggered field, and the ordering wave vector $Q=(1-m)\pi$,
where the magnetization density $m=M_z/M_s$,
which is the ratio of magnetization $M_z$ to its saturation value $M_s$.
In practice, the staggered field can be effectively induced from three-dimensional (3D) magnetic ordering of quasi-1D materials,
such as $\rm YbAlO_3$ \cite{wu_TLL_2019,Fan_phase_2020,nikitin_multiple_2021,Yang_CBS_2023},
$\rm SrCo_2V_2O_8$ \cite{Bera_Magnetic_2014}
and $\rm BaCo_2V_2O_8$ \cite{Zou_E8_2021}.
We note that the staggered field can be both commensurate and incommensurate, depending on whether $2\pi/Q$ is a rational or irrational number, respectively.

\begin{figure}[t]
    \centering
    \includegraphics[width=0.99\columnwidth]{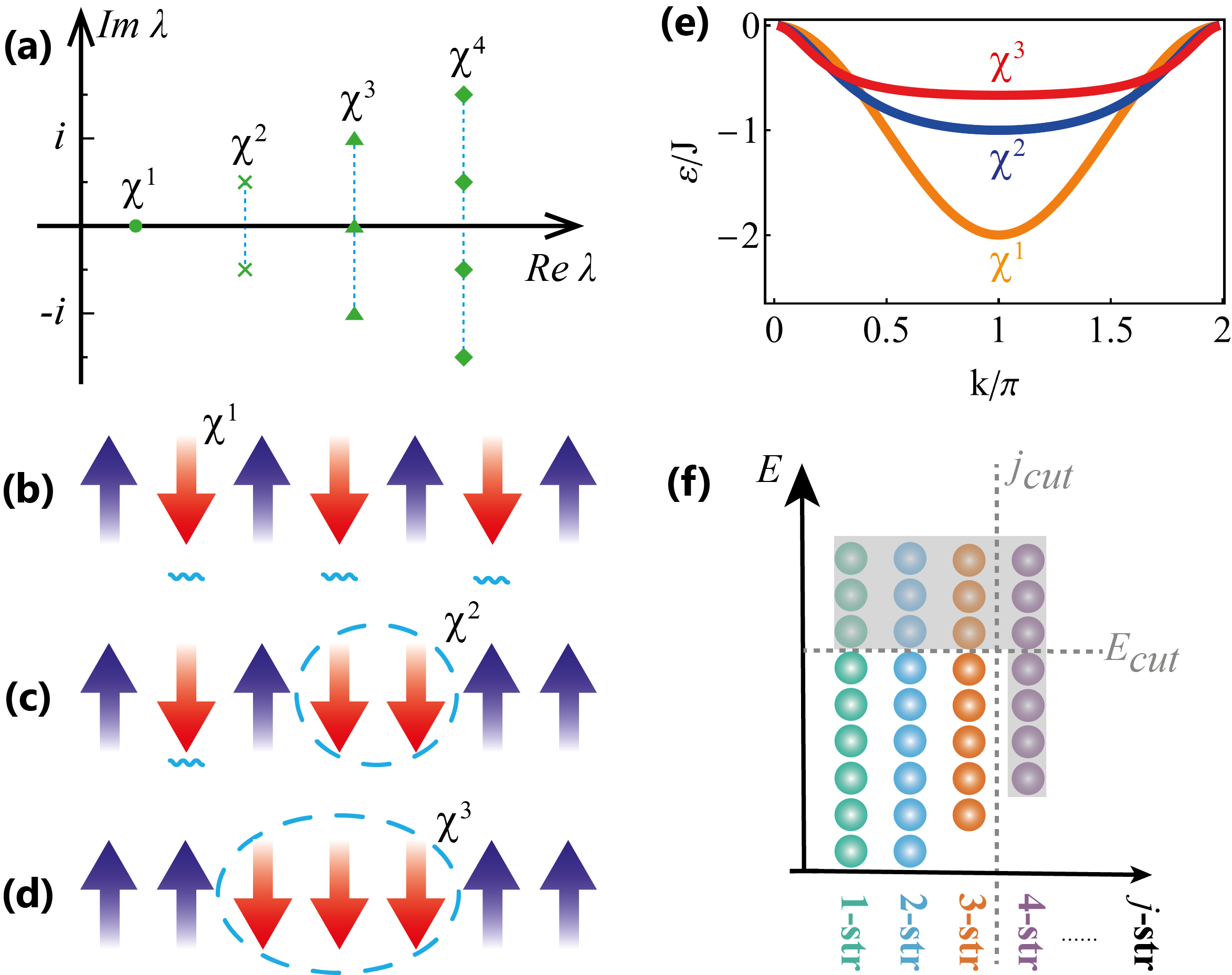}
    \caption{ (a) The rapidities of Bethe strings with different lengths in the complex plane.
    (b)-(d) Pictorial spin configurations of the Bethe strings.
    View the down spin (red arrows) and up spin (blue arrows) as the magnon and vacuum, respectively.
    A single Bethe string $\chi^j$ contains $j$ bounded magnons.
    (e) The energy-(quasi)momentum relation for Bethe strings $\chi^j$.
    (f) Illustration of the truncated string state space, where each ball denotes a $j$-string state.
    The shaded region is separated by vertical and horizontal lines, representing the cutoffs for string length and the energy, respectively.
    }
    \label{fig:raps_TS}
\end{figure}

\section{Exact Bethe string state}
\label{sec:Bethe string state}

In this section, we begin with an introduction to the coordinate Bethe ansatz and Bethe string states for the Hamiltonian $H_0$ [Eq.~\eqref{eq:H0}].
Then an efficient method is presented for obtaining the exact solutions from the BAE.

\subsection{Bethe ansatz and the Bethe string state}

Due to U(1) symmetry of $H_0$ [Eq.~\eqref{eq:H0}]
the magnetization $M_z=1/2-M/N$ is the conserved quantity,
where $M$ is the number of down spins,
i.e. magnons, with respect to
the fully polarized state with all up spins $|\uparrow\cdots\uparrow\rangle$.
In the coordinate Bethe ansatz \cite{bethe1931,karabach_introduction_1997,karbach_intro_2000},
the eigenstate of $H_0$ [Eq.~\eqref{eq:H0}] is
the Bethe state with $M$ magnons,
which is determined by a set of rapidities $\{\lambda_l\}_M$
satisfying the BAE,
\be
\left(
\frac{\lambda_l+i/2}{\lambda_l-i/2}
\right)^N
=
\prod_{k\neq l}
\left(
\frac{\lambda_l-\lambda_k+i}{\lambda_l-\lambda_k-i}
\right),
\label{eq:BAE}
\ee
with $l=1,\cdots,M$.
The corresponding quasi-momentum $k_l=\pi-2\arctan(2\lambda_l)$.
These rapidities $\{\lambda_l\}_M$ manifest as either complex-conjugate pairs or real numbers [Fig.~\ref{fig:raps_TS}(a)] \cite{vladimirov_proof_1986}.
The pair of complex rapidities implies significant physical property:
the corresponding magnons exhibit an intriguing phenomenon in coordinate space, forming effectively bounded magnons commonly known as ``Bethe string''
\cite{takahashi_one-dimensional_1971,karabach_introduction_1997,Takahashi1999}.
And the length of the string is determined by the number of rapidities with a common real center.
Intuitively, Bethe string $\chi^{j}$ ($j\geq2$) of length $j$ is a ``big'' quasiparticle
in which $j$ bounded magnons move coherently, referred to as a $j$-string [Fig.~\ref{fig:raps_TS}(b)-(d)].
When $j=1$, the 1-string $\chi^{1}$ is just the unbound magnon.
Correspondingly, the rapidities of a string $\chi^j$ takes the form \cite{takahashi_one-dimensional_1971,takahashi_one-dimensional_1972}
\be
\lambda^n_{j,\alpha} =
\lambda_{j,\alpha}+\frac{i}{2}(j+1-2n)+d^n_{j,\alpha},
\label{eq:complx_root}
\ee
where $n=1,\ldots,j$ denotes the $j$th magnon in the $j$-string.
The number of $j$-strings is denoted as $M_j$,
and $\alpha = 1,\ldots,M_j$ label different $j$-strings with the same length $j$.
Thus we have $\sum_j jM_j=M$ for an $M$-magnon Bethe state.
We refer to $\lambda_{j,\alpha}$ as
the string center which gives the real part of the $j$-string if the deviation $d^n_{j,\alpha}$ is omitted.
Under the assumption $d^n_{j,\alpha}=0$,
we obtain eigenenergy of a Bethe string state, $E=\sum_{j,\alpha}\varepsilon_{j,\alpha}$, with
$\varepsilon_{j,\alpha}=
{-2jJ}/{(4\lambda_{j,\alpha}^2+j^2)}$.
Therefore, we can show the relation between energy and quasi-momentum for different strings
in Fig.~\ref{fig:raps_TS}(e).
For finite \( d_{j,\alpha}^n \), the eigenenergy becomes $E=\sum_{j,\alpha,n}\varepsilon_{j,\alpha}^n$ with \(\varepsilon_{j,\alpha}^n = {-2J}/{[4(\lambda^n_{j,\alpha})^2 + 1]}\).

To ensure clarity in terminology, we refer to a Bethe state with all $M$ magnons being $\chi^1$ as the 1-string state.
For $j>1$, we classify an $n*j$-string state
with $M_j=n$ and $M_1=M-n*j$.
When $n=1$, it is simply referred to as the $j$-string state.
This convention can be consistently extended to cover other cases.
\begin{table*}
\begin{ruledtabular}
\begin{tabular}{cccccc}
 & $\lambda_3^1$ & $\lambda_3^2$ & $\lambda_3^3$ &
$\lambda_1^1$ & $\lambda_1^2$ \\
\hline
unphysical & $0.4955 + 0.9622i$ & $0.4955 - 0.9622i$ & 0.4458 & 0.4458 & 0.1803
\\
physical & $0.4918 + 0.9615i$ & $0.4918 + 0.9615i$ & $0.4448 + 0.0188i$ & $0.4448 - 0.0188i$ & 0.1807
\end{tabular}
\end{ruledtabular}
\caption{
The solutions of a 3-string state to Bethe ansatz equations Eq.~\eqref{eq:BAE} with $N=12$ and $M=5$.
The solutions presented in the first row are obtained with the method in Ref.~\cite{hagemans_deformed_2007},
while those in the second row are obtained with our method.
Note that in the first row, $\lambda_3^3$ and $\lambda_1^1$ coincide, indicating an unphysical outcome.
}
\label{tab:table}
\end{table*}

\subsection{Exact solution}
\label{sec:Exact solution}

To characterize Bethe string states, the initial step is to obtain rapidities $\{\lambda_j\}$ ($j=1,\ldots,M$) by solving the BAE [Eq.~\eqref{eq:BAE}].
This is commonly achieved by considering the logarithmic form of BAE,
\be
\frac{2\pi}{N}I_j=\Theta_1(\lambda_j)-\frac{1}{N}\sum_{k=1}^M\Theta_2(\lambda_j-\lambda_k),
\label{eq:lBAE}
\ee
where
$\Theta_j(\lambda)=2\arctan\left(2\lambda/j\right)$,
and $I_j$ is the corresponding Bethe quantum number.
Equation~\eqref{eq:lBAE} is highly efficient for finding the real solutions using the iterative method \cite{hagemans_deformed_2007,press_numerical_2007}.
However, for complex solutions, we first need to consider the reduced Bethe equation with $d_{j,\alpha}^n\to0$ \cite{Takahashi1999},
\be
\frac{2\pi}{N}
I_{j,\alpha}
=
\Theta_j(\lambda_{j,\alpha})
-\frac{1}{N}
\underset{(k,\beta)\neq(j,\alpha)}{\sum_{k=1}^M
\sum_{\beta=1}^{M_k}}
\Theta_{jk}(\lambda_{j,\alpha}-
\lambda_{k,\beta}),
\label{eq:rlBAE}
\ee
$\forall j$
with $M_j\neq 0$,
and $\alpha=1,\cdots,M_j$,
with
$\Theta_n(\lambda)=2\arctan\left({2\lambda}/n\right)$,
and
$\Theta_{nm}=
(1-\delta_{nm})
\Theta_{|n-m|}+
2\Theta_{|n-m|+2}
+\cdots+
2\Theta_{n+m-2}
+\Theta_{n+m}$.
The $I_{j,\alpha}$ is referred to as the reduced Bethe quantum number.
The Eq.~\eqref{eq:rlBAE} can also be tackled iteratively to obtain the string centers $\{\lambda_{j,\alpha}\}$,
whose associated complex solutions are constructed from Eq.~\eqref{eq:complx_root}
with $d_{j,\alpha}^n=0$.
Nevertheless, these solutions are generally not exact because they disregard the finite deviation $d_{j,\alpha}^n$.
By utilizing the solutions obtained with Eq.~\eqref{eq:rlBAE} as an initial guess,
the finite deviation is accessible for a majority of the string states following the method in Ref.~\cite{hagemans_deformed_2007}.
A summary is presented in Appendix.~\ref{app:exact solution}.

However, the strategy introduced above fails to generate all exact solutions
when the number of lattice sites exceeds $N\gtrsim12$,
with example in Table~\ref{tab:table}
and details in Appendix~\ref{app:exact solution}.
The limitations arise from the generation of repeated real rapidities in string states
which are physically forbidden \cite{deguchi_non_2015,com_rep_root}.
The key to solving the problem is that
the repeated real rapidities actually form a complex conjugate pair with a minor imaginary part (typically $\lesssim1/N$).
To implement the observation into the algorithm,
we divide and conquer,
with details in Appendix~\ref{app:repeated real rapidities}.
For instance, for a 3-string state, the typical rapidity pattern is that three
of them share a common real part
up to a finite deviation $d_{j,\alpha}^n$,
while the remaining rapidities are all real.
However, when we encounter repeated real rapidities (in first row of Table~\ref{tab:table}), usually involving the 3-string center and one real rapidity,
we introduce a small imaginary part to create a complex conjugate pair.
This pair and other rapidities are then treated as a new initial guess for the BAE, from which we are able to efficiently obtain the exact solution,
the second row in Table~\ref{tab:table}.

Before ending this section, it is imperative to underscore the importance of exact solutions.
As shown in the Appendix \ref{app:determinant formula},
the determinant expression of $\langle\mu|\sigma^{z}|\lambda\rangle$
becomes divergent when $|\lambda\rangle$ represents string states with zero deviation due to the failure of regularization.
Therefore, assuming $d^n_{j,\alpha}=0$ may not be appropriate for our subsequent TS${}^3$A approach.
Therefore, the practical route is to consider the string states with finite $d_{j,\alpha}^n$.

\begin{figure*}[t]
    \centering
    \includegraphics[width=0.29\textwidth]{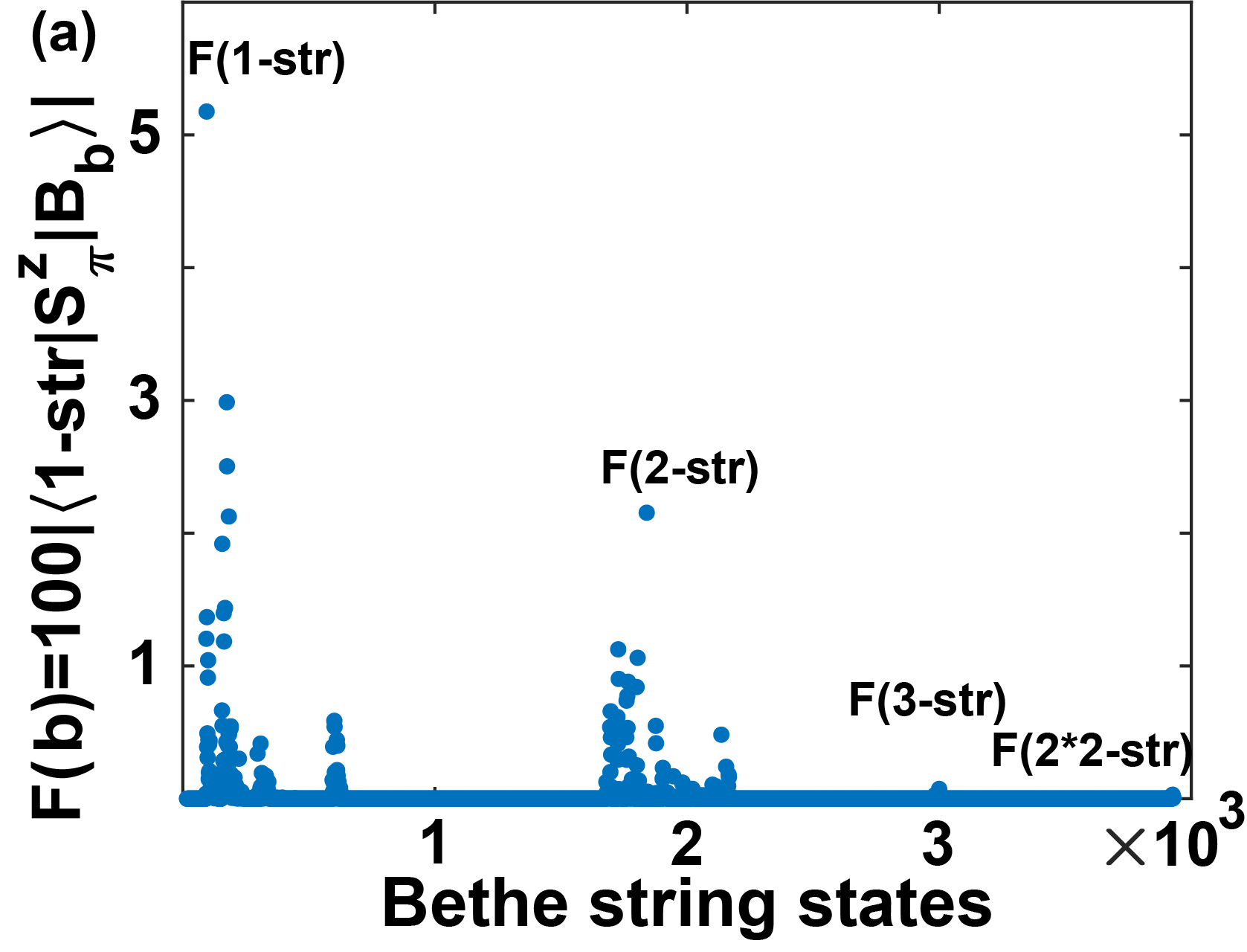}
    \includegraphics[width=0.3\textwidth]{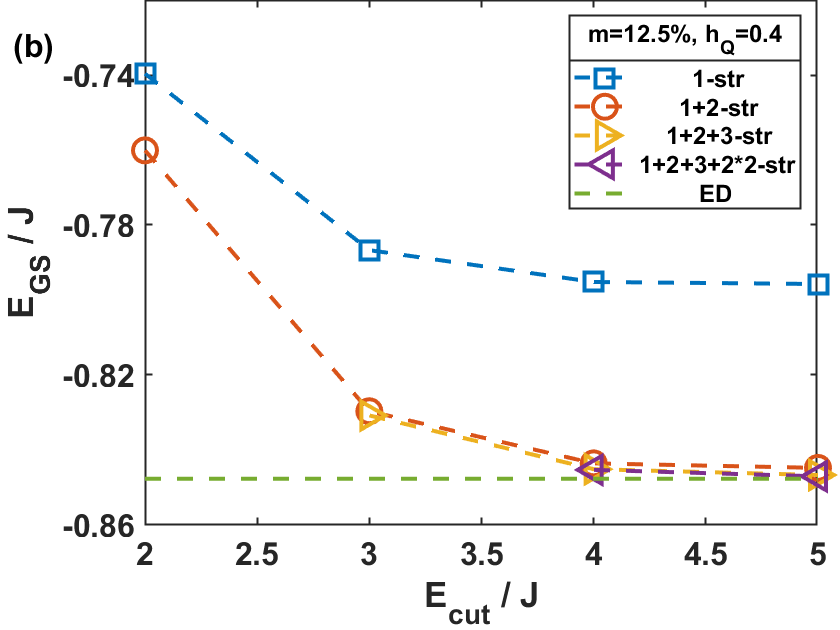}
    \includegraphics[width=0.3\textwidth]{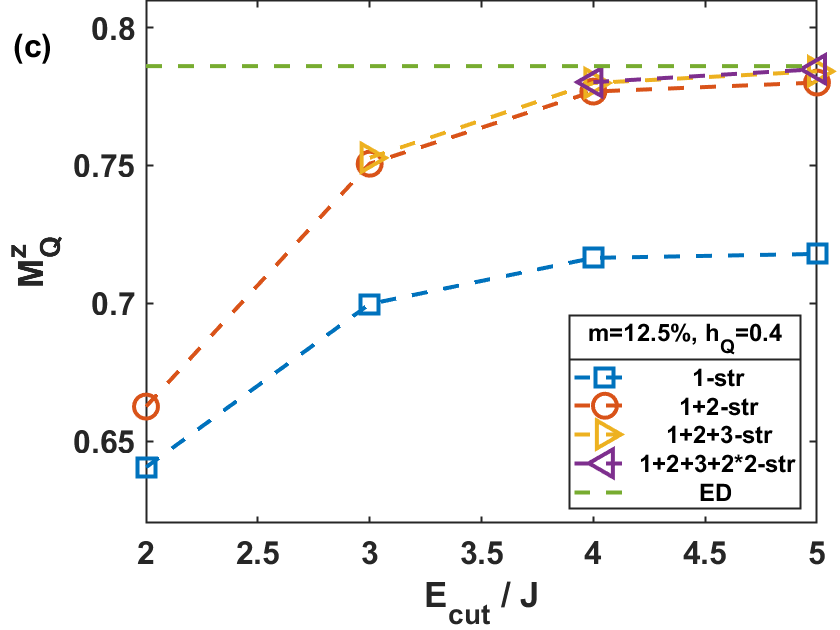}
    \caption{
    (a) The absolute value of matrix entry $\langle \mbox{1-str}|S_\pi^z|B_b\rangle$ where Bethe string state $|B_b\rangle$ ranges from 1-string to $2*2$-string states.
    (b) Ground state energy $E_{GS}$ and (c) staggered magnetization $M_Q^z$ calculated with different energy cutoffs and combinations of string states at $N=16$, depicted by dashed lines with symbols.
    Note that the gap of 3- and $2*2$-string states are $\simeq2.88J$ and $\simeq3.61J$, respectively.
    The green dashed line denotes the results from exact diagonalization.
    }
    \label{fig:FF_Sz}
\end{figure*}

\section{Truncated string-state-space approach}
\label{sec:TS3A}

In many physical problems,
our focus is on the low-energy subspace rather than the entire Hilbert space \cite{james_non-perturbative_2018,yurov_truncated_1990,Caux_Constructing_2012}.
In this study, we employ the TS${}^3$A method, as illustrated in [Fig.~\ref{fig:raps_TS}f],
to gain insights into the low-energy physics of nonintegrable Heisenberg models Eq.~\eqref{eq:H}.
The detailed construction is as follows.

When considering a nonintegrable perturbation, such as $H'$ [Eq.~\eqref{eq:Hprime}],
the Bethe string state is no longer the eigenstate.
Therefore, it becomes necessary to find out the new ground state and low-energy excited states before conducting any calculations of physical quantities, such as correlation functions.
The first step is to construct the matrix representation of the nonintegrable Hamiltonian $H$ [Eq.~\eqref{eq:H}]
within a truncated low-energy subspace of Bethe string states,
denoted as
$H^{tr}_{ab}=\delta_{ab}E^B_a+\langle B_a|H'| B_b \rangle$,
with $E^B_a \leq E^B_{cut}$.
The truncated dimension of $H^{tr}_{ab}$ is typically much less than $2^N$, which is determined by the energy cutoff $E^B_{cut}$ and types of Bethe string states $| B_a \rangle$.
Following the diagonalization of $H^{tr}$, a new ground state $|GS\rangle$ and low-energy excited states are obtained, which are considered as approximate eigenstates of the Hamiltonian $H$.

However, an immediate question arises: How do we select the types of Bethe string states and determine the energy cutoff?
To answer the first question, we investigate the form factor between Bethe string states.
It is evident from Fig.~\ref{fig:FF_Sz}(a) that the form factors for string states generally diminish rapidly as the difference in string length increases.
For instance, the ground state of $H_0$ Eq.~\eqref{eq:H0} is a 1-string state, and then we can safely truncate the string state space into relevant subspace in terms of string lengths.
For the second issue,
we study the asymptotic behavior of ground state energy $E_{GS}$ and the staggered magnetization
$M^z_Q=\sum_j e^{-iQj} \langle GS|S_j^z|GS\rangle/\sqrt{N}$
as the energy cutoff of the truncated space increase.
In Fig.~\ref{fig:FF_Sz}(b,c),
the calculation includes all allowed string states within a given $E_{cut}$.
As $E_{cut}$ increases the results converge rapidly and
approach the exact values obtained from the ED calculation.
Notably, even if only 1- and 2-string states are considered, the obtained results 
are already very close to the exact values,
while the impact of 3- and $2*2$-string states is marginal.
This phenomenon confirms the suggestion that
the relevant string states primarily arise from those with small length differences,
as illustrated in Fig.~\ref{fig:FF_Sz}(a).

\section{Spin dynamics}
\label{sec:spin dynamics}

To investigate nonintegrable spin dynamics of $H$ [Eq.~\eqref{eq:H}],
we focus on the zero-temperature dynamical structure factor (DSF)
for spin along longitudinal ($z$) direction ($\hbar=1$),
\be
D^{zz}(q,\omega)=2\pi
\sum_{\mu}
|\langle GS|S_q^z|\mu \rangle|^2
\delta(\omega-E_\mu+E_{GS}),
\label{eq:DSF_no_T}
\ee
with $q$ being the transfer momentum and $\omega$ being the transfer energy between the ground state $|GS\rangle$ and excited states $|\mu\rangle$, whose energies are $E_{GS}$ and $E_\mu$, respectively.
In the following calculation, the eigenstates
$|GS\rangle$ and $|\mu\rangle$ are obtain by the TS$^3$A developed in Sec.~\ref{sec:TS3A}.
The form factor $\langle GS|S_q^z|\mu \rangle$ is deeply related to the form factor of Bethe string states,
which can be elegantly expressed in terms of determinant
\cite{kohno_string_dynamically_2009,caux_four-spinon_2006,caux_2008,caux_two-spinon_2008,castillo_exact_2020,yang_string_1D_2019},
with a summary in Appendix~\ref{app:determinant formula}.

\begin{figure*}
    \centering
    \includegraphics[width=0.9\textwidth]{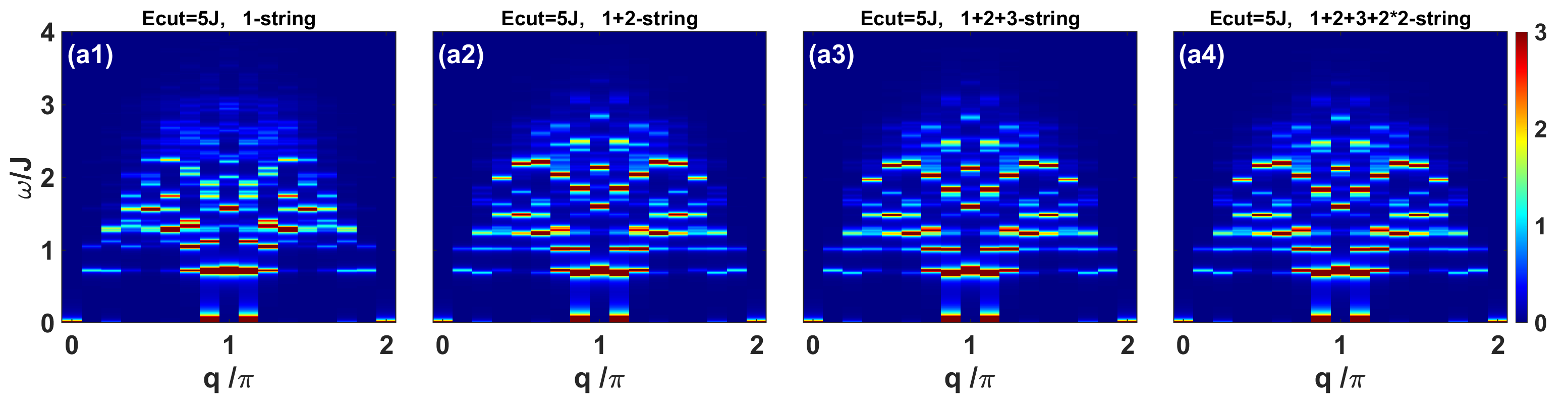}
    \\
    \includegraphics[width=0.9\textwidth]{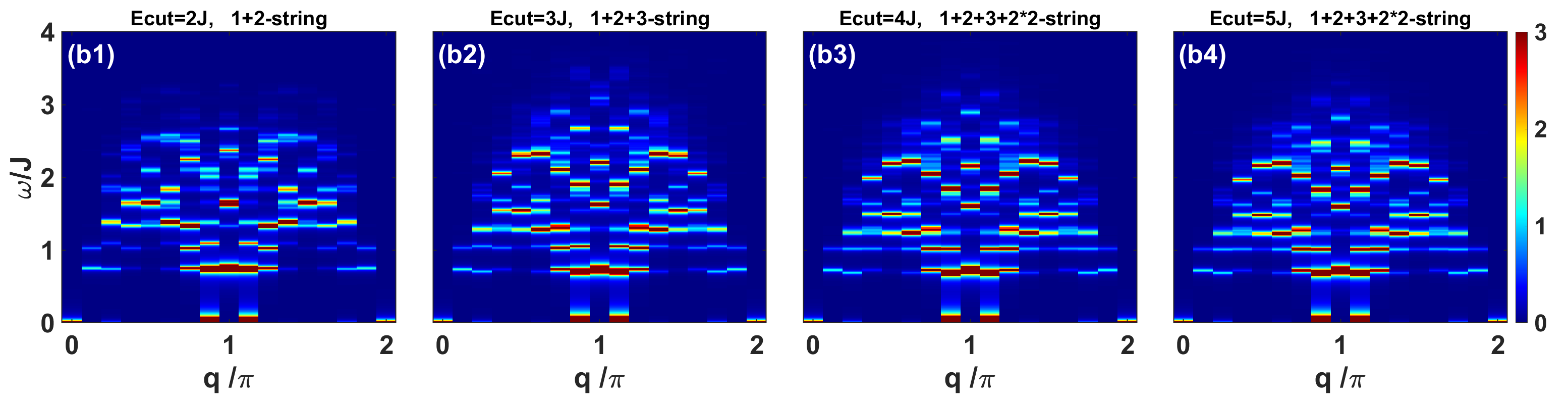}
    \caption{
    Zero temperature DSF $D^{zz}$ at $N=16$, $m=12.5\%$, $h_Q=0.4J$ under different truncation schemes:
    (a1-a4) Different selected string types with fixed energy cutoff $E_{cut}=5J$;
    (b1-b4) Different energy cutoffs with 1-, 2-, 3-, and $2*2$-string states.
    Note that the gap of 3 and $2*2$ string states are $\simeq2.88J$ and $\simeq3.61J$, respectively.
    The $\delta$ function in the DSF is broadened via a Lorentzian
    function $\frac{1}{\pi}\gamma/[(\omega-E_\mu+E_{GS})+\gamma^2]$ with $\gamma$=0.02.
    }
    \label{fig:TS3A_TruncationSchemes}
\end{figure*}

\begin{figure*}
    \centering
    \includegraphics[width=0.45\textwidth]{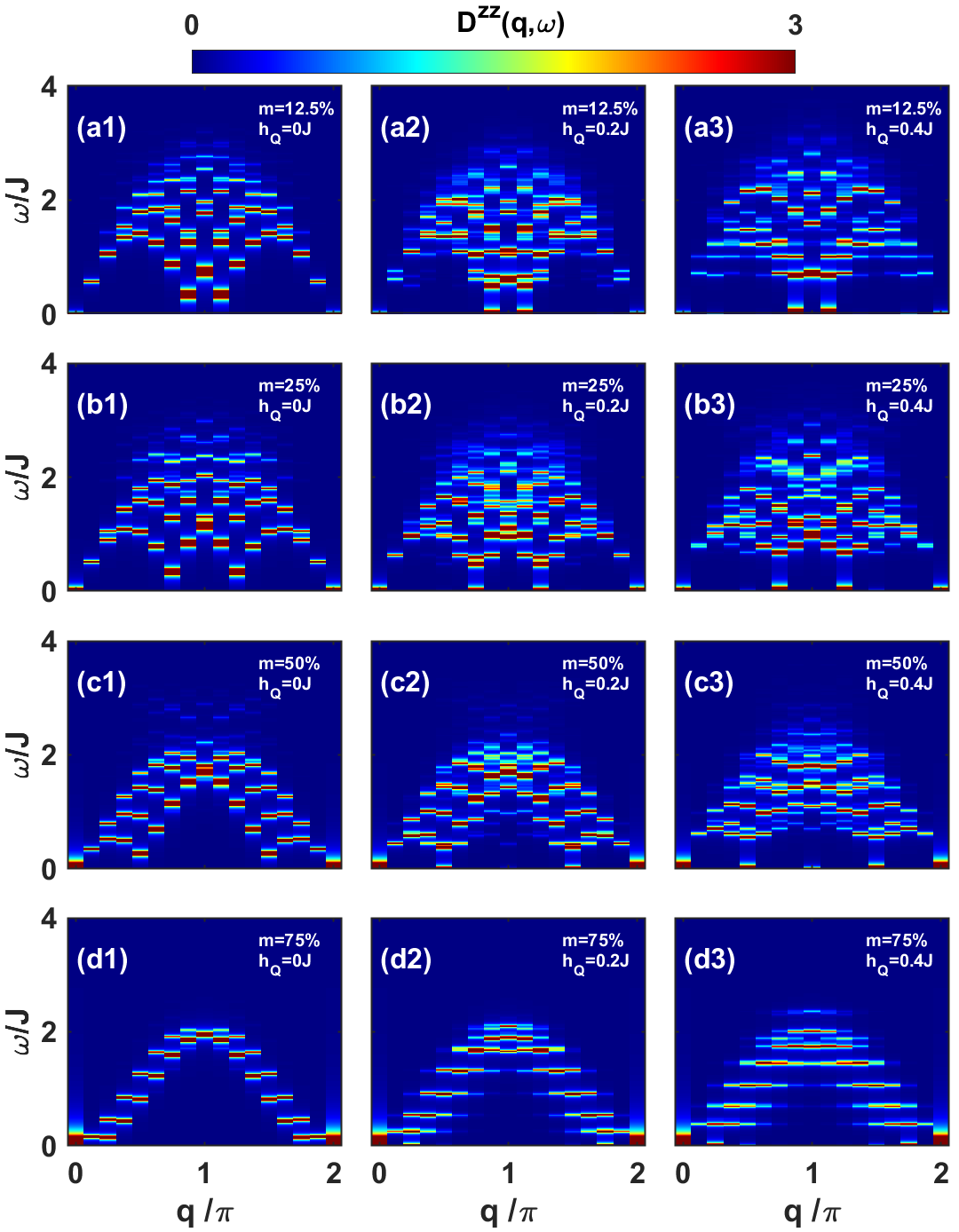}
    \includegraphics[width=0.45\textwidth]{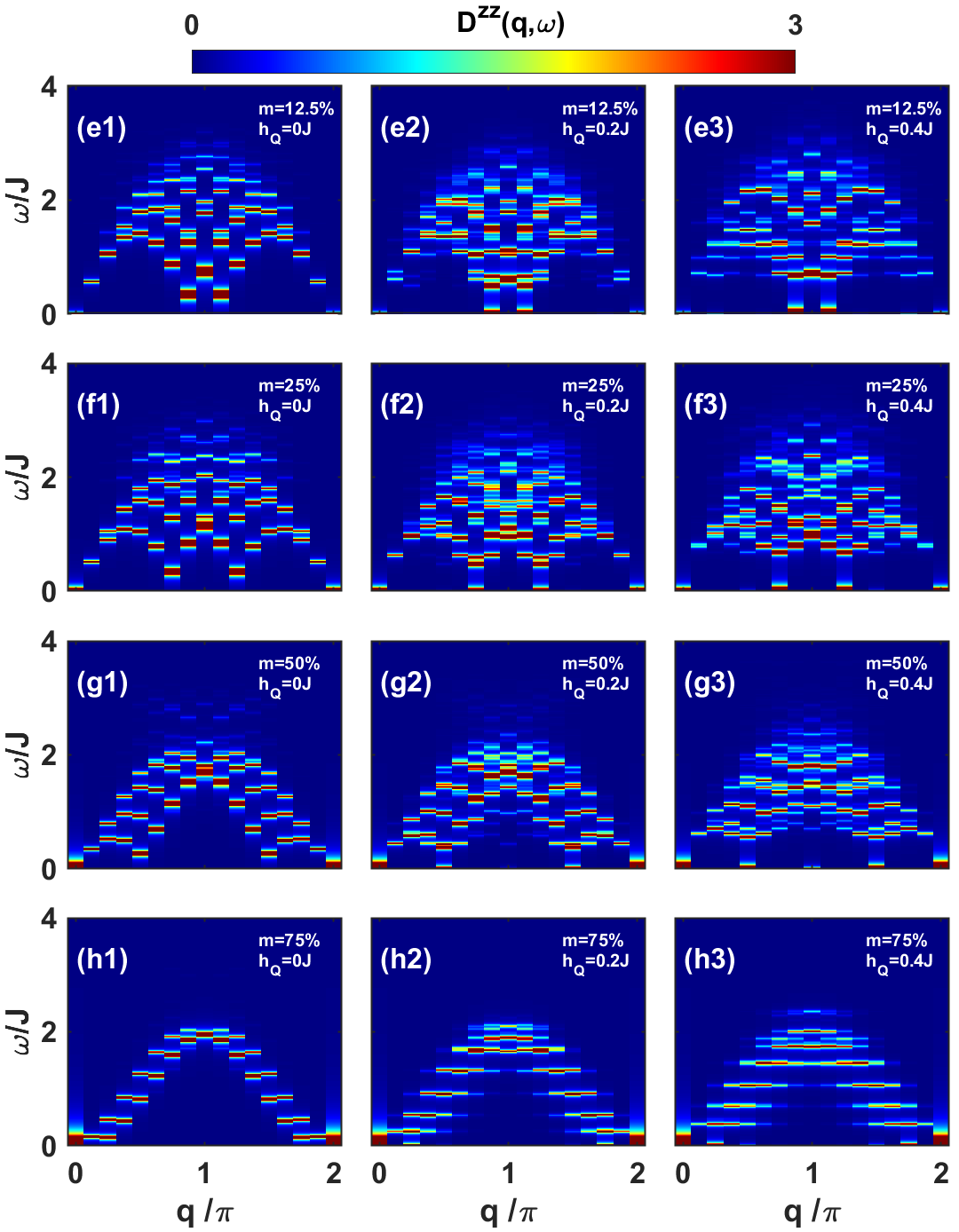}
    \caption{
    Zero temperature DSF $D^{zz}(q,\omega)$ with different magnetization $m$ and staggered field $h_Q$ with lattice site $N=16$.
    (a1-d3) The results obtained from TS$^3$A consider the string types including 1, 2, 3, and $2*2$ string states with
    fixed energy cutoff $E_{cut}=5J$.
    (e1-h3) ED results with the same lattice size.
    The $\delta$ function in the DSF is broadened via a Lorentzian
    function $\frac{1}{\pi}\gamma/[(\omega-E_\mu+E_{GS})+\gamma^2]$ with $\gamma$=0.02.
    }
    \label{fig:ComparisonDSFs}
\end{figure*}

To begin with,
we investigate the TS$^3$A results under different truncation schemes for $N=16$, $m=12.5\%$, and $h_Q=0.4J$.
In Fig.~\ref{fig:TS3A_TruncationSchemes}(a1-a4),
the DSF is calculated with different selected string types and fixed energy cutoff $E_{cut}=5J$,
showing that 1- and 2-strings are the dominant states in the spin dynamics.
In Fig.~\ref{fig:TS3A_TruncationSchemes}(b1-b4),
the DSF is calculated with different energy cutoff and fixed string types (including 1-, 2-, 3- and $2*2$-string),
showing that the dynamical spectrum converges quickly as $E_{cut}$ increases.
Furthermore,
we compare the DSF results at $N=16$ obtained from TS$^3$A to that of the ED method \cite{LinHQ_exact_1990,jung_guide_2020},
whose results reveal a remarkable agreement in Fig.~\ref{fig:ComparisonDSFs}.
This excellent comparison suggests that
the TS$^3$A is a highly efficient method for studying the nonintegrable spin dynamics.

\begin{figure}[t]
\centering
\includegraphics[width=0.85\columnwidth]{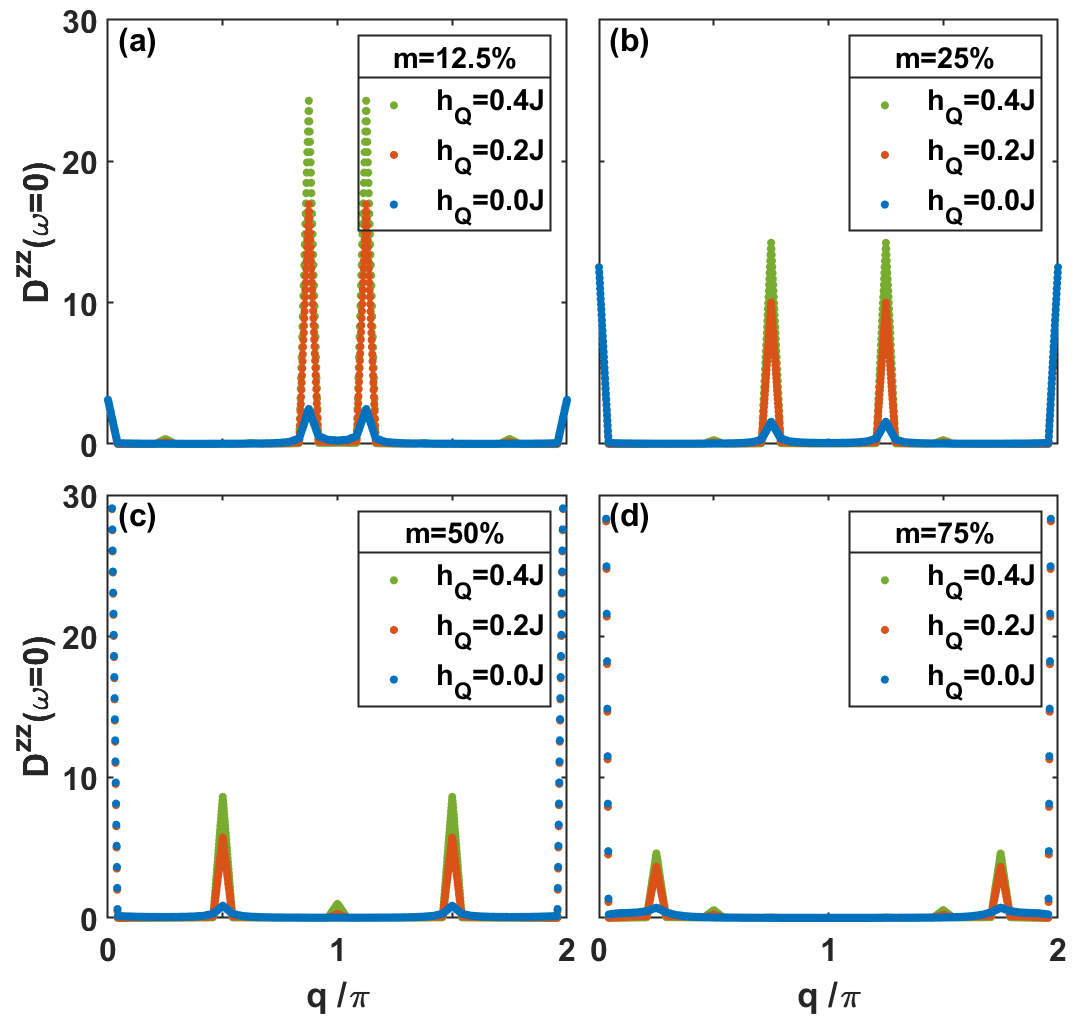}
\\
\includegraphics[width=0.8\columnwidth]{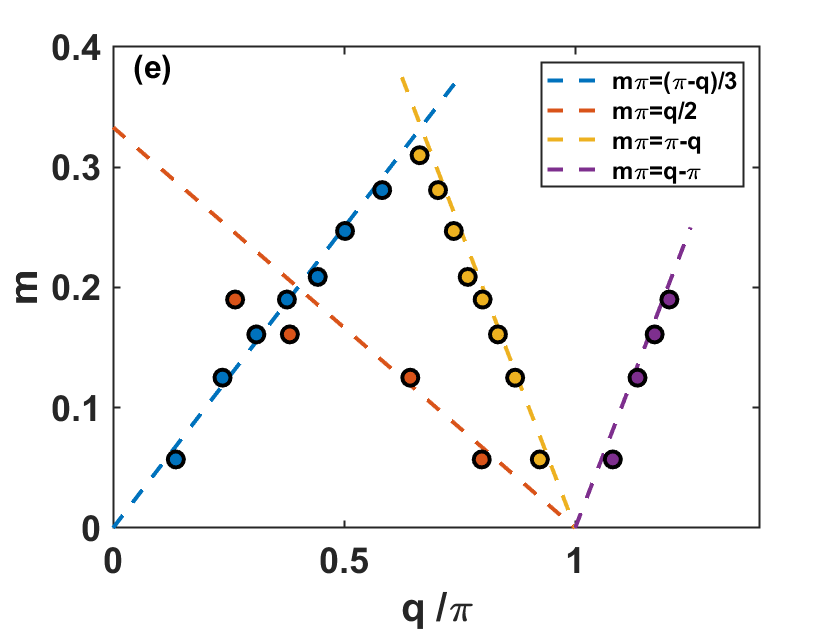}
\caption{(a-d) The static structure factor $D^{zz}(\omega=0)$ with $N=48$, $h_{Q}=0J$, $0.2J$, and $0.4J$, 
and magnetization density $m$=12.5\%, 25\%, 50\%, 75\%, obtained from the TS$^3$A.
(e)
Comparison between experimental data (filled circles) and theoretical predictions (dashed lines).
The filled circles represent satellite peaks obtained from quasi-elastic neutron scattering data, as extracted from Ref.~\cite{nikitin_multiple_2021}.
The dashed lines represent
the theoretical prediction for the elastic peaks.
}
\label{fig:satellites}
\end{figure}

In the following, we present the DSF results obtained from the TS$^3$A at $N=48$.
Due to the staggered field perturbation term $H'$ [Eq.~\eqref{eq:Hprime}],
where $\sum_n\cos(Qr_n)S^z_n \propto (S^z_{Q}+S^z_{-Q})$,
the non-vanishing matrix element of $H'$ only appears between Bethe states with momentum difference $\Delta q = \pm Q$.
As a result, the ground state consists of Bethe states with momenta $nQ\ (|n|=0, 1, 2, \ldots)$.
For static structure factor $D^{zz}(q,\omega=0)$,
a series of staggered-field-induced peaks
appear at $nQ$ ($|n|=1, 2, \ldots$),
manifesting the presence of multi-$Q$ Bethe states in the ground state.
For instance, in Fig.~\ref{fig:satellites}(a-d), there are satellite peaks at $q=2Q,-2Q+2\pi$ in addition to the predominant peaks at $q=Q,-Q+2\pi$, with $Q=(1-m)\pi$.

\begin{figure}[t]
    \centering
    \includegraphics[width=0.99\columnwidth]{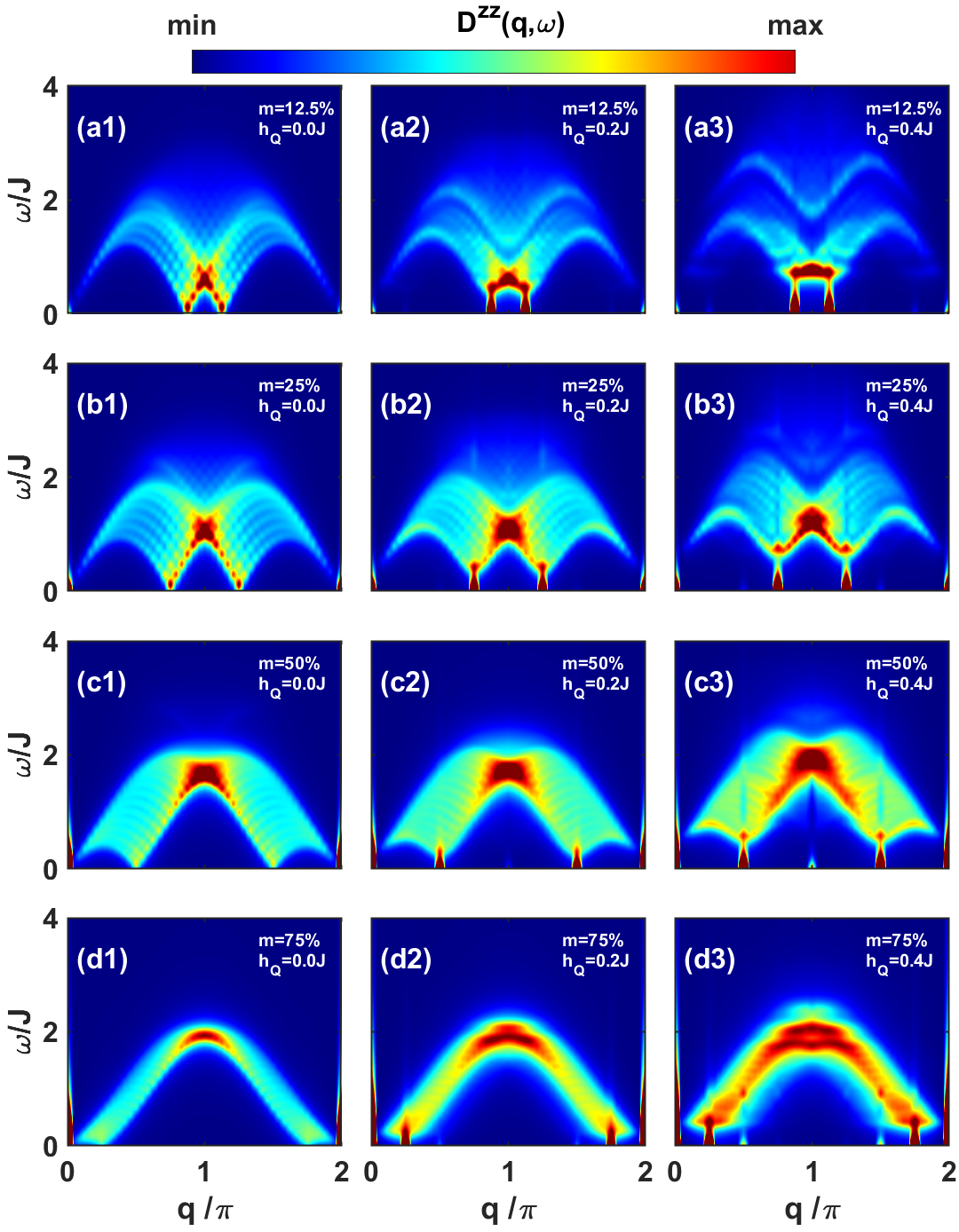}
    \caption{Zero temperature DSF $D^{zz}(q,\omega)$ with $N=48$ obtained from the TS$^3$A. The DSFs with {$h_{Q}=0J$, $0.2J$, and $0.4J$} (from left to right), and magnetization density $m$=12.5\%, 25\%, 50\%, 75\% (from top to bottom).
    The $\delta$ function in the DSF is broadened via a Lorentzian
    function $\frac{1}{\pi}\gamma/[(\omega-E_\mu+E_{GS})+\gamma^2]$ with $\gamma$=0.12.
    The data are further interpolated along the horizontal direction for 800 query points with equal spacing.
    }
    \label{fig:DSF_IC}
\end{figure}

In Fig.~\ref{fig:DSF_IC}, when $h_{Q}=0$,
the dynamical spectra exhibit gapless excitations at $q=(1\pm m)\pi$,
and the 2-string states are barely separated from the broad continuum of 1-string states.
When $h_{Q}>0$, an energy gap emerges both near the elastic line and between the continua of 1- and 2-string states.
This phenomenon arises because the staggered field acts as a confining field for the Heisenberg spin chains, effectively restricting the motion of spins
\cite{lake_spinon_confinement_2010,bera_spinon_2017,wang_confined_2016,Essler_q1DHeisenberg_1997,Starykh_q1DSDW_2014}.
These induced gaps reflect the energy cost for the excitation of Bethe strings, which is known as the confinement of Bethe strings.
At small magnetization [Fig.~\ref{fig:DSF_IC}(a1-a3)], 2-string states are effectively confined and become separated from the 1-string continuum.
However,
as magnetization $m$ increases, the 2-string continuum gradually dissipates into higher energy ranges.

Notably, the capability for larger size calculation of TS$^3$A not only reduces the finite size effect
but also renders the characteristics of spectra more transparent and discernible.
And, the TS${}^3$A offers two key advantages compared to the ED method:
First, it has higher efficiency, facilitating much larger systems ($N\gtrsim50$); second,
it naturally provides a unified Bethe-string-based physical picture for understanding the underlying physics.

We conclude this section by emphasizing that our model and findings offer a direct application in comprehending the low-energy spin dynamics observed in the quasi-1D antiferromagnet $\rm YbAlO_3$ \cite{wu_TLL_2019,nikitin_multiple_2021,Yang_CBS_2023}.
In this material, the low-energy effective Hamiltonian is described by the 1D Heisenberg models Eq.~\eqref{eq:H0}, and Eq.~\eqref{eq:H} if the material is 3D ordered.
In Fig.~\ref{fig:satellites}(e),
the quasi-elastic signals obtained from neutron scattering align with the theoretical predictions,
providing compelling evidence for the coexistence of multi-$Q$ Bethe states in the ordered phase of $\rm YbAlO_3$.
Moreover, the staggered field, arising from the 3D ordering,
plays the role of a confining field coupled with the spin chains within the material.
As a result, the distinctive features characterizing the confined string states are observed through the inelastic neutron scattering spectra of $\rm YbAlO_3$
 \cite{Yang_CBS_2023}.

\section{Conclusion}
\label{sec:conclusion}

We exploited an efficient routine to find exact solutions for the Bethe string states from the BAE of the spin-$\frac{1}{2}$ Heisenberg spin chain.
Based on the exact solutions
we further developed the TS${}^3$A
which enabled us to determine eigenstates and eigenenergies of nonintegrable spin-$\frac{1}{2}$ Heisenberg systems with U(1) symmetry preserved.
The method was then applied to systematically study the spin dynamics of the spin-$\frac{1}{2}$ Heisenberg spin chain under staggered field.
In the dynamical spectra, we revealed a series of elastic peaks located at the integer multiples of the ordering wavevector $Q$,
signifying the existence of multi-$Q$ Bethe string states within the ground state.
Moreover, the staggered field serves as a confining field for Bethe string states,
inducing confinement gaps between the continua of 1- and 2-string states.

Our TS${}^3$A
machine offers a Bethe-string-based scenario, contributing to a more 
comprehensive understanding of Heisenberg spin systems.
We have demonstrated the efficiency and validity of this framework by interpreting experimental observations of the quasi-1D antiferromagnet $\rm YbAlO_3$.
This intriguing consistency between theoretical predictions based on the TS$^3$A and experimental results motivates its extended application to ladder and two-dimensional Heisenberg systems.
This broadening of scope not only enhances the versatility of the Bethe string picture but also transcends its conventional one-dimensional limitations.

\section*{Acknowledgments}
We thank Yunfeng Jiang for helpful discussion.
This work is supported by National Natural Science Foundation of China No. 12274288
and the Innovation Program for Quantum Science and Technology Grant No. 2021ZD0301900,
and the Natural Science Foundation of Shanghai with
grant No. 20ZR1428400.


\appendix


\section{Iterative method for exact solution}
\label{app:exact solution}

This appendix present the iterative method for solving the Bethe equation.
Note that it's sufficient to solve the highest weight state containing only finite rapidities,
while other states can be obtained by adding infinite rapidities \cite{hagemans_deformed_2007}.

\subsection{Deviation \texorpdfstring{$d_{j,\alpha}^n=0$}{}}
\label{app:d=0}

For the 1-string state, all rapidities are real, which can be directly solved from the iterative form of the Bethe equation,
\be
\lambda_j = \frac{1}{2}
\tan\left[ \frac{\pi}{N}I_{j}
+\frac{1}{N}\sum_{k=1}^M
\arctan(\lambda_j-\lambda_k)
\right]
\ee
where $\{I_{j}\}$ is the corresponding Bethe quantum number for $\{\lambda_j\}$.

For the string state, there is at least one complex rapidity in the pattern of Eq.~\eqref{eq:complx_root}.
To obtain the corresponding rapidities,
we convert the reduced Bethe equation Eq.~\eqref{eq:rlBAE} into the iterative form,
\be
\lambda_{j,\alpha}
=
\frac{j}{2}
\tan
\left[
\frac{\pi}{N}
I_{j,\alpha}
+
\frac{1}{2N}
\underset{(k,\beta)\neq(j,\alpha)}{\sum_{k=1}^M
\sum_{\beta=1}^{M_k}}
\Theta_{jk}(\lambda_{j,\alpha}-
\lambda_{k,\beta})
\right],
\ee
where $\{I_{j,\alpha}\}$ is the corresponding reduced Bethe quantum number for string centers $\{\lambda_{j,\alpha}\}$.
Following Eq.~\eqref{eq:complx_root},
the complex string states is constructed from $\{\lambda_{j,\alpha}\}$ with $d_{j,\alpha}^n=0$.

\subsection{Deviation \texorpdfstring{$d_{j,\alpha}^n\neq0$}{}}
\label{app:dneq0}

To determine the exact deviation $\{d_{j,\alpha}^n\}$, the strategy becomes more intricate for the XXX model \cite{hagemans_deformed_2007} and for the gapped XXZ model \cite{yang_string_1D_2019}.
Here, we only consider 2- and 3-string states for illustration.

For a string with length $j=2$,
its two complex rapidities are $\lambda_j^{+,-}=\lambda_j^{1,2}=
\lambda_j^0\pm\frac{i}{2}+ d_j^{1,2}$,
where the deviations are purely imaginary,
$d_{j}^1=i\delta^1_j$
and
$d_{j}^2=i\delta^2_j=-i\delta^1_j$.
Then we can have the first-order deviation,
\be
\delta_{j=2}^1\approx
\left( \frac{\lambda_j^+-i/2}{\lambda_j^++i/2} \right)^{N}
\left(
\prod_{k}^{real}
\frac{\lambda_j^+-\lambda_k+i}{\lambda_j^+-\lambda_k-i}
\right).
\ee
Next, utilizing the first-order deviation,
we can determine the true Bethe quantum number $J^{1,2}$ from reduced one $I_2$,
\be
J^1=J^2 - \Theta_H(\delta)=\frac{1}{2}
\left( I_2 + \frac{N}{2} \mbox{sign}(\lambda_j^0) - \Theta_H(\delta) \right)
\ee
where
\be
\Theta_H(\delta)=
\frac{N}{2}-M+1
+I_2,
\quad
\mbox{mod 2}.
\label{eq:def2_sgn_delta}
\ee
Then, considering the sum of the logarithmic Bethe equations Eq.~\eqref{eq:lBAE},
\be
\sum_{\sigma\in \{+,-\}}\Theta_1(\lambda^{\sigma})
=\frac{1}{N}\sum_{\sigma\in \{+,-\}}
\left(
2\pi I_\sigma+\sum_{k=1}^M\Theta_2(\lambda^\sigma-\lambda_k)
\right),
\label{eq:def2_sum_lBAE}
\ee
and
the deformation of Bethe equation Eq.~\eqref{eq:BAE},
\be
\frac{\lambda^+-\lambda^--i}{\lambda^+-\lambda^-+i}
=
\left(
\frac{\lambda^+-i/2}{\lambda^++i/2}
\right)^N
\prod_k
\frac{\lambda^+-\lambda_k+i}{\lambda^+-\lambda_k-i},
\label{eq:def2_BAE}
\ee
we can solve for $\lambda^\pm$, along with $\{\lambda_1\}_{M-2}$ from the Bethe equations Eq.~\eqref{eq:lBAE} of 1-strings.

For a string with length $j=3$,
it contains three rapidities,
$\lambda_j^0$, $\lambda_j^{+,-}=\lambda_j^{1,2}=
\lambda_j^0\pm i + d_j^{1,2}$,
where $d_j^{1}=(d_j^{2})^*=\epsilon^1+i\delta^1$.
Then we can have the first-order deviation,
\be
 d_{j=3}^1\approx
6i\cdot
\left(
\frac{\lambda_j^1+i/2}{\lambda_j^1-i/2}
\right)^{-N}
\left(
\prod_{k}^{real}
\frac{\lambda_j^1-\lambda_k+i}{\lambda_j^1-\lambda_k-i}
\right).
\ee
We note that for the 3-string, $\mbox{Im}(\lambda^+)>1/2$ must hold,
which leads to the fact that $J^\pm$ must be a wide pair with
$J^- - J^+ = 1$.
Then, we still need two more equations to solve the true Bethe quantum numbers $J^0$, $J^\pm$.
The first equation is the sum of the logarithmic Bethe equations Eq.~\eqref{eq:lBAE},
\be
\begin{split}
J^+ + J^0 +J^-
&=
I_3-\frac{1}{2}\sum_{k=1}^{\text{1-str}}
\mbox{sign}(\lambda-\lambda_{k}).
\end{split}
\label{eq:def3_BQN_pm0}
\ee
Another necessary equation is the sum of logarithmic BAEs of $\lambda^\pm$
\be
\begin{split}
&2\pi(J^++J^-)
+\Theta_2(\lambda^+ - \lambda^0)
+\Theta_2(\lambda^- - \lambda^0)
\\&\quad =
N(\Theta_1(\lambda^+)+\Theta_1(\lambda^-))
-\Theta_2(\lambda^+ - \lambda^-)
-\Theta_2(\lambda^- - \lambda^+)
\\&\quad
-\sum_{k=1,\beta}(\Theta_2(\lambda^+-\lambda_{k,\beta})
+\Theta_2(\lambda^--\lambda_{k,\beta})),
\end{split}
\label{eq:def3_BAE_pm}
\ee
Let A be the right hand side of Eq.~\eqref{eq:def3_BAE_pm}.
Because $\Theta_2(\lambda^+ - \lambda^0)
+\Theta_2(\lambda^- - \lambda^0)\in (-2\pi,2\pi)$,
$J^+ +J^-$ is the
even(odd) integer number in $(A/2\pi-1,A/2\pi+1)$
when $M$ is even(odd).
Therefore,
\be
\begin{split}
J^++J^-
&=
(1+(-1)^M)
\left\lfloor
\frac{1}{2}\left(
\frac{A}{2\pi}+1
\right)
\right\rfloor
\\&\quad
+
(1-(-1)^M)\left(
\left\lfloor
\frac{1}{2}\left(
\frac{A}{2\pi}+1
\right)+\frac{1}{2}
\right\rfloor
-\frac{1}{2}
\right).
\end{split}
\label{eq:def3_BQN_pm}
\ee
Now combing the wide pair condition ($J^- - J^+ = 1$), Eqs.~\eqref{eq:def3_BQN_pm0}, and \eqref{eq:def3_BQN_pm},
$J^\pm$ and $J^0$  can be determined.
The Bethe quantum number $\{J_k \}$ for real rapidities can be shown to be of the following expression,
\be
J_k
=I_k
-\frac{1}{2}
\mbox{sgn}(\lambda_{k}-\lambda^0_{j=3}).
\ee
To solve rapidities, we first need the sum of logarithmic BAE of $J^\pm$ and $J^0$ without setting $\epsilon$ and $\delta$ to be zero
\be
\begin{split}
&\frac{2\pi}{N}
(J^++J^0+J^-)
=
\Theta_1(\lambda^+)+\Theta_1(\lambda^0)+\Theta_1(\lambda^-)
\\&
-\frac{1}{N}
\sum_{k}^{real}
\Theta_2(\lambda^+-\lambda_{k})
+\Theta_2(\lambda^0-\lambda_{k})
+\Theta_2(\lambda^--\lambda_{k}).
\end{split}
\label{eq:def3_BAE_pm0}
\ee
The second equation is the sum of logarithmic BAE of $J^\pm$ Eq.~\eqref{eq:def3_BAE_pm}.
The third one is obtained from Bethe equation Eq.~\eqref{eq:BAE} after some simple manipulation
\be
\begin{split}
\frac{(\lambda^+-\lambda^0)-i}{(\lambda^+-\lambda^0)+i}
&=
\frac{(\lambda^+-\lambda^-)+i}{(\lambda^+-\lambda^-)-i}
\cdot
\prod_k
\frac{(\lambda^+-\lambda_k)+i}{(\lambda^+-\lambda_k)-i}
\\&\quad\cdot
\left(
\frac{\lambda^+-i/2}{\lambda^++i/2}
\right)^N.
\end{split}
\label{eq:def3_BAE}
\ee
The logarithmic Bethe equations Eq.~\eqref{eq:lBAE} are also needed for real rapidities.

\begin{table*}[t]
\centering
\begin{ruledtabular}
\begin{tabular}{ccccc}
 & $\{J\}_{M=5}$ & $\{I\}_{M=5}$ & $\{\lambda\}_{M=5}$ &
$Energy$ \\
\hline
1. unphysical & 4 &         & 0.495521913637784 + 0.962224932131036i & -3.632275481625215 \\
           & 3 & $1_3$   & {0.445792844757107 + 0.000000000000000i} & \\
           & 5 &         & 0.495521913637784 - 0.962224932131036i & \\
           & 2 & $3/2_1$ & 0.180317318693691 + 0.000000000000000i & \\
           & 3 & $5/2_1$ & {0.445792844757134 + 0.000000000000000i} & \\

\\
2. physical& 4 &         & 0.491814213695900 + 0.961471132379077i & -3.60069325626932 \\
           & 3 & $1_3$   & {0.444763506448628 + 0.018770199402376i} & \\
           & 5 &         & 0.491814213695898 - 0.961471132379085i & \\
           & 2 & $3/2_1$ & 0.180714318631831 + 0.000000000000000i & \\
           & 3 & $5/2_1$ & {0.444763506448649 - 0.018770199402378i} &  \\
\end{tabular}
\end{ruledtabular}
\caption{
The solutions of a 3-string state to Bethe ansatz equations Eq.~\eqref{eq:BAE} with $N=12$ and $M=5$.
The unphysical solutions with repeated rapidities are obtained from the method described in Appendix~\ref{app:d=0} and \ref{app:dneq0}.
The physical solutions are obtained from the method described in Appendix~\ref{app:repeated real rapidities}.
}
\label{tab:app N12_M5_3strs}
\end{table*}

Here, we present the 3-string state results of $N=12$ and $M=5$ obtained from the above iterative method
in the first set of Table~\ref{tab:app N12_M5_3strs}.
We can observe that two real rapidities coincide, one is the string center and another is a 1-string.
However, it is unphysical because of the incorrect eigenenergy and the absence of wave function under this set of 
solutions.

\subsection{Repeated real rapidities}
\label{app:repeated real rapidities}

To tackle the issue of the repeated real rapidities of $\lambda^0$,
we introduce a small imaginary part to create a complex conjugate pair, as required by the BAE.
Now, we have two complex conjugate pairs.
The first pair has a small imaginary part, $\lambda^{0\pm}=\lambda_0\pm i\delta_0$,
while the second one has a larger imaginary part around $\pm i$,
$\lambda^{3\pm}=\lambda_3 \pm i(1+\delta_3)$.
Note that 4 complex rapidities need 4 equations to solve.
The strategy is similar to the procedures mentioned above.
Two equations come from the sum of logarithmic Bethe equations of $\lambda^{0\pm}$ and $\lambda^{3\pm}$.
Another two equations come from the original Bethe equations of $\lambda^{0+}$ and $\lambda^{3+}$.
Combining the logarithmic Bethe equations for real rapidity,
we could solve $\lambda^{0\pm}$, $\lambda^{3\pm}$,
and real rapidities $\{\lambda_1\}_{M-4}$.

Then, we re-determine the 3-string state for $N = 12$ and $M = 5$ [in the 2$^{nd}$ set of Table~\ref{tab:app N12_M5_3strs}].
Now, this set of rapidities is the exact solution of the original Bethe equation Eq.~\eqref{eq:BAE},
which is consistent with Ref.~\cite{deguchi_non_2015},
rational $Q$-system method and ED calculation.

\section{The determinant formula}
\label{app:determinant formula}

\subsection{Norm of the Bethe state}

Given a set of rapidities ${\lambda_j}$ ($j=1,\ldots,M$) representing exact solutions of the Bethe Ansatz equation Eq.~\eqref{eq:BAE}, the norm of the corresponding Bethe state is expressed as
\cite{Franchini2017,hagemans_deformed_2007,slavnov_calculation_1989},
\be
\begin{split}
\mathbb{N}_M(\{\lambda_j\})&=
(-1)^{M}
\frac{\prod_{j\neq k}
(\lambda_j-\lambda_k+i)}
{\prod_{j\neq k}
(\lambda_j-\lambda_k)}
\det\Phi(\{\lambda\}),
\end{split}
\label{eq:norm}
\ee
where the matrix elements of $\Phi$ are
\be
\begin{split}
\Phi_{ab}&=
\delta_{ab}\left[N\frac{4}{1+4\lambda_a^2}
-\sum_k
\frac{2}{1+(\lambda_a-\lambda_k)^2}
\right]
\\&
+(1-\delta_{ab})
\frac{2}{1+(\lambda_a-\lambda_b)^2}
\end{split}
\ee

\subsection{Form factors}
\label{app:FF}

The non-zero form factors associated with $\sigma^z$ correspond to states characterized by equal magnon numbers,
\be
\begin{split}
F_j^z(\{\mu\}_M,\{\lambda\}_M)&=
\langle \{\mu\}_M| \sigma_j^z | \{\lambda\}_M \rangle
\\&=
\frac{\phi_{j-1}(\{\mu\}_M)}{\phi_{j-1}(\{\lambda\}_M)}
\prod_{l=1}^M\frac{(\mu_l+i/2)}{(\lambda_l+i/2)}
\\&\cdot
\frac{i^{M}\det(H-2P)}
{\prod_{l>m}(\mu_l-\mu_m)\prod_{l<m}(\lambda_l-\lambda_m)}.
\end{split}
\ee
where $\phi_j(\{\lambda\}_M)=e^{-iq_\lambda r_j}$, and $q_\lambda$
is the eigen-momentum of Bethe state $|\{\lambda\}_M\rangle$.
The matrix elements of $H$ and $P$ matrix are defined as
\be
\begin{split}
H_{ab}&=
\frac{1}{(\mu_a-\lambda_b)}
\left(
\prod^{M}_{l\neq a}
(\mu_l-\lambda_b+i)
\right.\\&\left.
-\left( \frac{\lambda_b-i/2}{\lambda_b+i/2} \right)^N
\prod^{M}_{l\neq a} (\mu_l-\lambda_b-i)
\right),
\end{split}
\label{eq:Hmatrix}
\ee
\be
P_{ab}=\frac{\prod_{l=1}^M(\lambda_l-\lambda_b+i)}{(\mu_a+i/2)(\mu_a-i/2)},
\label{eq:Pmatrix}
\ee
respectively.
Here we note that both the 2- and 3-string states with $d_{j,\alpha}^n=0$ can cause divergence in the $P$ matrix Eq.~\eqref{eq:Pmatrix}.
However, the divergence can not be regularized since there are no common 
terms in $H$ matrix Eq.~\eqref{eq:Hmatrix}.

\;

\;

\;

\;

\;

\;

\bibstyle{apsrev-nourl}
\bibliography{Refs_TS3}

\begin{thebibliography}{81}
\bibitem [{com()}]{com_rep_root}%
  \BibitemOpen
\bibfield  {number} {  }\href@noop {} {}\bibinfo {howpublished} {In the
  spin-1/2 XXX chain, we can check this invalidity by inserting the unphysical
  solutions into the wavefunction.}\BibitemShut {Stop}%
\end{thebibliography}

\end{document}